# Implementation of Kalman Filter with Python Language

**Mohamed LAARAIEDH**
*IETR Labs, University of Rennes 1*
Mohamed.laaraiedh@univ-rennes1.fr

## Abstract

In this paper, we investigate the implementation of a Python code for a Kalman Filter using the Numpy package. A Kalman Filtering is carried out in two steps: Prediction and Update. Each step is investigated and coded as a function with matrix input and output. These different functions are explained and an example of a Kalman Filter application for the localization of mobile in wireless networks is given.

## I. Introduction

Within the significant toolbox of mathematical tools that can be used for stochastic estimation from noisy sensor measurements, one of the most well-known and often-used tools is what is known as the Kalman filter. The Kalman filter is named after Rudolph E. Kalman, who in 1960 published his famous paper describing a recursive solution to the discrete-data linear filtering problem [3].

The Kalman filter is essentially a set of mathematical equations that implement a predictor-corrector type estimator that is optimal in the sense that it minimizes the estimated error covariance when some presumed conditions are met. Since the time of its introduction, the Kalman filter has been the subject of extensive research and application, particularly in the area of autonomous or assisted navigation. This is likely due in large part to advances in digital computing that made the use of the filter practical, but also to the relative simplicity and robust nature of the filter itself. A complete tutorial about Kalman filtering is given in [2].

### I.1. Mathematical Formulation of Kalman Filter

The Kalman filter addresses the general problem of trying to estimate the state $x \in \Re^n$ of a discrete-time controlled process that is governed by the linear stochastic difference equation

$$x_k = A x_{k-1} + B u_k + w_{k-1} \tag{1}$$

with a measurement $y \in \Re^m$ that is

$$y_k = H x_k + v_k \tag{2}$$

The random variables $w_k$ and $v_k$ represent the process and measurement noise respectively. They are assumed to be independent of each other, white, and with normal probability distributions

$$p(w) \approx N(0, Q) \tag{3}$$
$$p(v) \approx N(0, R) \tag{4}$$

In practice, the process noise covariance $Q$ and measurement noise $R$ covariance matrices might change with each time step or measurement, however here we assume they are constant. The $n \times n$ matrix $A$ relates the state at the previous time step to the state at the current step, in the absence of either a driving function or process noise. The $n \times 1$ matrix $B$ relates the optional control input $u \in \Re^l$ to the state x. The $m \times n$ matrix $H$ in the measurement equation relates the state to the measurement $y_k$. For more deep investigation of the Kalman Filter you may see the reference [2].

## I.2. How does Kalman Filter work?

The Kalman filter process has two steps: the prediction step, where the next state of the system is predicted given the previous measurements, and the update step, where the current state of the system is estimated given the measurement at that time step. The steps translate to equations as follows [2]:

- Prediction

$$X_k^- = A_{k-1} X_{k-1} + B_k U_k \tag{5}$$

$$P_k^- = A_{k-1} P_{k-1} A_{k-1}^T + Q_{k-1} \tag{6}$$

- Update

$$V_k = Y_k - H_k X_k^- \tag{7}$$

$$S_k = H_k P_k^- H_k^T + R_k \tag{8}$$

$$K_k = P_k^- H_k^T S_k^{-1} \tag{9}$$

$$X_k = X_k^- + K_k V_k \tag{10}$$

$$P_k = P_k^- - K_k S_k K_k^T \tag{11}$$

where
• $X_k^-$ and $P_k^-$ are the predicted mean and covariance of the state, respectively, on the time step $k$ before seeing the measurement.
• $X_k$ and $P_k$ are the estimated mean and covariance of the state, respectively, on time step $k$ after seeing the measurement.
• $Y_k$ is mean of the measurement on time step $k$.
• $V_k$ is the innovation or the measurement residual on time step $k$.
• $S_k$ is the measurement prediction covariance on the time step $k$.
• $K_k$ is the filter gain, which tells how much the predictions should be corrected on time step $k$.

## I.3. Applications of the Kalman Filter

The applications of the Kalman Filtering in real world are diverse. An example application would be providing accurate, continuously updated information about the position and velocity of an object given only a sequence of observations about its position, each of which includes some error. For a similar, more concrete example, in a radar application, where one is interested in tracking a target, information about the location, speed, and acceleration of the target is measured at each time instant, with a great deal of degradation by noise. The Kalman filter exploits the dynamics of the target, which govern its time evolution, to remove the effects of the noise and get a good estimate of the location of the target at the present time (filtering), at a future time (prediction), or at a time in the past (interpolation or smoothing). Other applications are weather forecasting, speech enhancement, economics, autopilot... etc.

## II. Python Code of the Kalman Filter

We have chosen to divide the Kalman Filtering Code in two parts similarly to its mathematical theory. The code is simple and divided in three functions with matrix input and output.

### II.1. Prediction Step

This step has to predict the mean $X$ and the covariance $P$ of the system state at the time step $k$. The Python function **kf_predict** performs the prediction of these output ($X$ and $P$) when giving six input:

$X$ : The mean state estimate of the previous step ($k-1$).
$P$ : The state covariance of previous step ($k-1$).
$A$ : The transition $n \times n$ matrix.
$Q$ : The process noise covariance matrix.
$B$ : The input effect matrix.
$U$ : The control input.

The Python code of this step is given by:

```
from numpy import dot
def kf_predict(X, P, A, Q, B, U):
    X = dot(A, X) + dot(B, U)
    P = dot(A, dot(P, A.T)) + Q
    return(X,P)
```

### II.2. Update Step

At the time step $k$, this update step computes the posterior mean $X$ and covariance $P$ of the system state given a new measurement $Y$. The Python function **kf_update** performs the update of $X$ and $P$ giving the predicted $X$ and $P$ matrices, the measurement vector $Y$, the measurement matrix $H$ and the measurement covariance matrix $R$. The additional input will be:

$K$ : the Kalman Gain matrix
$IM$ : the Mean of predictive distribution of $Y$
$IS$ : the Covariance or predictive mean of $Y$
$LH$ : the Predictive probability (likelihood) of measurement which is computed using the Python function **gauss_pdf**.

The Python code of these two functions is given by:

```
from numpy import dot, sum, tile, linalg
from numpy.linalg import inv
```

```python
def kf_update(X, P, Y, H, R):
    IM = dot(H, X)
    IS = R + dot(H, dot(P, H.T))
    K = dot(P, dot(H.T, inv(IS)))
    X = X + dot(K, (Y-IM))
    P = P - dot(K, dot(IS, K.T))
    LH = gauss_pdf(Y, IM, IS)
    return (X,P,K,IM,IS,LH)

def gauss_pdf(X, M, S):
    if M.shape()[1] == 1:
        DX = X - tile(M, X.shape()[1])
        E = 0.5 * sum(DX * (dot(inv(S), DX)), axis=0)
        E = E + 0.5 * M.shape()[0] * log(2 * pi) + 0.5 * log(det(S))
        P = exp(-E)
    elif X.shape()[1] == 1:
        DX = tile(X, M.shape()[1])- M
        E = 0.5 * sum(DX * (dot(inv(S), DX)), axis=0)
        E = E + 0.5 * M.shape()[0] * log(2 * pi) + 0.5 * log(det(S))
        P = exp(-E)
    else:
        DX = X-M
        E = 0.5 * dot(DX.T, dot(inv(S), DX))
        E = E + 0.5 * M.shape()[0] * log(2 * pi) + 0.5 * log(det(S))
        P = exp(-E)

    return (P[0],E[0])
```

## III.  Example of application: Tracking of mobile in wireless network

The most interesting field of application of Kalman Filter, in telecommunications, is the tracking of a mobile user connected to a wireless network. In this section, we will present a simple tracking algorithm of a mobile user who is moving in a room and connected to at least three wireless antennas [1].

The matrix of measurement $Y$ describes the estimated position of the mobile using a trilateration algorithm based on a least square estimation and the knowledge of at least three values of Time of Arrival (ToA) at time step $k$. These values are computed using ranging procedures between the mobile and the three antennas [1].

Starting by an initialization of different matrices and using the updated matrices for each step and iteration, we plot in Fig- 1 the estimated, the real trajectory of the mobile user, and the measurements performed by the least square based trilateration. We show here that the Kalman Filter enhances the accuracy of tracking compared to the static least square based estimation. The Python code describing the tracking process is given as below. In order to simplify the understanding of this code, we draw the matrix $Y$ randomly centered on the true value of mobile position.

```
from numpy import *
from numpy.linalg import inv

#time step of mobile movement
dt = 0.1

# Initialization of state matrices
X = array([[0.0], [0.0], [0.1], [0.1]])
P = diag((0.01, 0.01, 0.01, 0.01))
A = array([[1, 0, dt , 0], [0, 1, 0, dt], [0, 0, 1, 0], [0, 0, 0,\
 1]])
Q = eye(X.shape()[0])
B = eye(X.shape()[0])
U = zeros((X.shape()[0],1))
```

```
# Measurement matrices
Y = array([[X[0,0] + abs(randn(1)[0])], [X[1,0] +\
 abs(randn(1)[0])]])
H = array([[1, 0, 0, 0], [0, 1, 0, 0]])
R = eye(Y.shape()[0])

# Number of iterations in Kalman Filter
N_iter = 50

# Applying the Kalman Filter
for i in arange(0, N_iter):
      (X, P) = kf_predict(X, P, A, Q, B, U)
      (X, P, K, IM, IS, LH) = kf_update(X, P, Y, H, R)
      Y = array([[X[0,0] + abs(0.1 * randn(1)[0])],[X[1, 0] +\
       abs(0.1 * randn(1)[0])]])
```

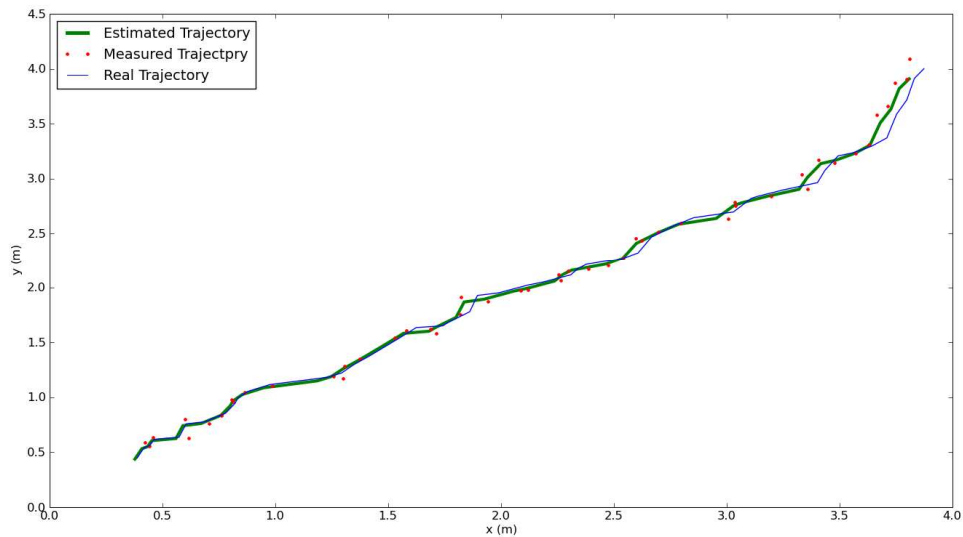

*Fig- 1: Kalman Filter Applied to ToA Based Localization*

## IV. Conclusions and future work

In this paper, we presented the Python code for the Kalman Filter implementation. We presented a two step based implementation and we give an example of using this kind of filters for localization in wireless networks. The next steps will be the implementation of others Bayesian filters like Extended Kalman Filter, Unscented Kalman Filter and Particle Filter. A third step of smoothing of estimations may be introduced later.